\documentclass[prl,twocolumn,letterpaper, superscriptaddress]{revtex4-1}
\usepackage{graphicx}
\usepackage{dcolumn}
\usepackage{bm}
\usepackage{color}
\usepackage{tabularx}
\usepackage{array}
\usepackage{amsmath}
\usepackage{stmaryrd}  

\begin{document}

\title{Magnonic spontaneous oscillation induced by parametric pumping}

\author{Yi Li}
\email{yili@anl.gov}
\affiliation{Materials Science Division, Argonne National Laboratory, Lemont, IL 60439, USA}

\author{Carissa Kiehl}
\affiliation{Materials Science Division, Argonne National Laboratory, Lemont, IL 60439, USA}
\affiliation{Physics and Astronomy Department, Carthage College, Kenosha, WI 53140, USA}

\author{Jinho Lim}
\affiliation{Department of Materials Science and Engineering, The Grainger College of Engineering, University of Illinois Urbana-Champaign, Urbana, IL, 61801, USA}
\affiliation{Materials Research Laboratory, The Grainger College of Engineering, University of Illinois Urbana-Champaign, Urbana, IL 61801, USA}

\author{Cliff Abbott}
\affiliation{Department of Physics and Energy Science, University of Colorado Colorado Springs, Colorado Springs, Colorado 80918, USA}

\author{Pratap K. Pal}
\affiliation{Materials Science Division, Argonne National Laboratory, Lemont, IL 60439, USA}
\affiliation{Department of Materials Science and Engineering, The Grainger College of Engineering, University of Illinois Urbana-Champaign, Urbana, IL, 61801, USA}

\author{Alex J. Szymczak}
\affiliation{Materials Science Division, Argonne National Laboratory, Lemont, IL 60439, USA}
\affiliation{Department of Physics, The Grainger College of Engineering, University of Illinois Urbana-Champaign, Urbana, IL 61820, USA}

\author{Juliang Li}
\affiliation{High Energy Physics Division, Argonne National Laboratory, Lemont, Illinois 60439, USA}

\author{Ralu Divan}
\affiliation{Center for Nanoscale Materials, Argonne National Laboratory, Argonne, IL 60439, USA}

\author{Clarence L. Chang}
\affiliation{High Energy Physics Division, Argonne National Laboratory, Lemont, Illinois 60439, USA}

\author{Charudatta Phatak}
\affiliation{Materials Science Division, Argonne National Laboratory, Lemont, IL 60439, USA}

\author{Dmytro A. Bozhko}
\affiliation{Department of Physics and Energy Science, University of Colorado Colorado Springs, Colorado Springs, Colorado 80918, USA}

\author{Axel Hoffmann}
\email{axelh@illinois.edu}
\affiliation{Department of Materials Science and Engineering, The Grainger College of Engineering, University of Illinois Urbana-Champaign, Urbana, IL, 61801, USA}
\affiliation{Materials Research Laboratory, The Grainger College of Engineering, University of Illinois Urbana-Champaign, Urbana, IL 61801, USA}
\affiliation{Department of Physics, The Grainger College of Engineering, University of Illinois Urbana-Champaign, Urbana, IL 61820, USA}

\author{Valentine Novosad}
\email{novosad@anl.gov}
\affiliation{Materials Science Division, Argonne National Laboratory, Lemont, IL 60439, USA}

\date{\today}

\begin{abstract}

Spontaneous dynamic systems have attracted significant attention for their rich underlying physics such as phase-locking and synchronization. In this work, we report a new mechanism of generating magnetic spontaneous oscillation via parametric pumping. By applying a pump tone to excite propagating spin waves in a yttrium iron garnet delay line, four-wave mixing converts the pump mode into two phase-autonomous propagating magnon modes, i.e. a spontaneous mode with nearly twice the wavenumber of the pump mode and an idler mode with nearly zero wavenumber. This allows us to reliably generate ultrasharp spin wave dynamics with broad frequency tunability from the pump and magnetic field. We show that the spontaneous mode can be phase-locked to a probe tone, similar to an auto-oscillator. Furthermore, the spontaneous dynamics can be used to implement a high-gain magnonic parametric amplifier with a gain up to 40 dB. Our results open a new avenue for studying nonlinear magnonics and synchronization physics in propagating magnon geometry and for developing new magnonic devices.

\end{abstract}

\maketitle

\section{Introduction}

Self-sustained spontaneous oscillation is defined as periodic oscillation without external drive, and is fundamental in nature, from the rhythmic beating of the heart to the synchronized flashing of fireflies.
Without phase reference, spontaneous oscillations are usually sensitive to phase perturbations and are subject to novel physical interactions, such as phase locking and synchronization \cite{Adler46,Kuramoto75,AcebronRMP05}. Particularly, this effect has been cultivated in spintronics through spin torque oscillators \cite{KiselevNature03,Kaka05, Mancoff05,HoussameddineNM07,PribiagNPhys07,SlavinIEEE09,SafranskiNC17}, where spin-polarized current injection compensates the magnetic damping and generates spontaneous magnetization dynamics at microwave frequencies, holding great promise in next-generation microelectronics \cite{ZhuIEEE08,Dieny20}, wireless communication \cite{SharmaNComm21}, neuromorphic computing \cite{Torrejon17,ZahedinejadNNano2020}, and unconventional computing \cite{Romera18,AlbertssonAPL21,HoushangPRApplied22}.

Besides damping compensation, another key mechanism for inducing spontaneous oscillation is parametric excitation. In this process, an external pump at a different frequency injects energy into the system, which is transferred to the oscillation mode through nonlinear interactions within the medium. Parametric excitation has been exploited for developing Josephson parametric oscillators (JPO) and optical parametric oscillators (OPOs), both of which play crucial roles in modern quantum and optical technologies. In magnonics, parametric excitation has inspired many novel discoveries such as magnon Bose-Einstein condensation \cite{DemokritovNature06,BozhkoNPhys16,SchneiderNNano2020,LvovPRL23}, parametric magnon amplification \cite{BaoAPL08,BracherAPL14,KamimakiPRApplied20,SchultheissSciAdv24} and magnon confluence process \cite{RomeroPRB09,QuPRB23,Makiuchi24,QuPRB25}. The process is primarily achieved through Suhl instability \cite{AndersonPR55, Suhl57} in terms of three-wave mixing (3WM) \cite{Kurebayashi11,LendinezNComm23,korberNComm23,NikolaevNComm24} and four-wave mixing (4WM) \cite{SchultheissPRB12,PirroPRL14} process. The former usually converts one pump magnon at frequency $f$ to two magnons at $f/2$; the latter converts two pump magnons at $f$ to two different magnons at $f_1$ and $f_2$ with $2f=f_1+f_2$. The above 3WM process only generates magnons which are strictly phase-bounded to the pump, meaning that they lack the phase adaptability to external stimulus. 4WM generates free-running magnons which are phase-independent from the pump. Nevertheless, the generated magnons usually span over a broad frequency range similar to thermal magnons \cite{SchultheissPRB12}, or follow irregular wavenumber-selecting process \cite{AnNComm24,GeilenAPR25}, leading to lack of controllability and reproducibility for applications.

In this work, we demonstrate robust excitation of magnon spontaneous oscillations in a yttrium iron garnet (YIG) thin-film delay line. By controlling the pump power to be slightly above the Suhl instability threshold, we trigger a 4WM process which converts the pump spin wave with wavenumber $k$ to a higher-frequency spin wave mode with nearly twice the wavenumber (spontaneous mode) and a lower-frequency spin wave mode with nearly zero wavenumber (idler mode). The spontaneous mode can be electrically detection when it is within the antenna-defined magnon transmission band, and can be generated over a broad frequency range by tuning the pump frequency and the external magnetic field. The idler mode cannot be always measured due to its decoupling with the antenna. We show the existence of the idler mode when the pump frequency is close to the Kittel mode. In addition, we show that the 4WM-generated spontaneous mode can be phase-locked to a probe signal. This proves that the spontaneous mode is phase-autonomous, i.e. it can adapt its phase to external stimulus without being bound to the pump. The idler mode can be phase-locked as well, and a parametric phase-locking of the spontaneous mode is simultaneously observed with the same locking bandwidth, proving their correlation in the 4WM process. Furthermore, we show that at weak probe powers well below the phase-locking regime, the spontaneous mode can act as a magnon parametric amplifier to a weak probe signal with a constant gain that is determined by the amplitude of the oscillator. The results open a new avenue of creating robust and highly controllable nonlinear spin wave dynamics in solid state, and utilizing the nonlinear dynamics in synchronization physics and microwave information processing.

\section{Results}

\subsection{Magnon spontaneous mode}

\begin{figure*}[htb]
 \centering
 \includegraphics[width=6.0 in]{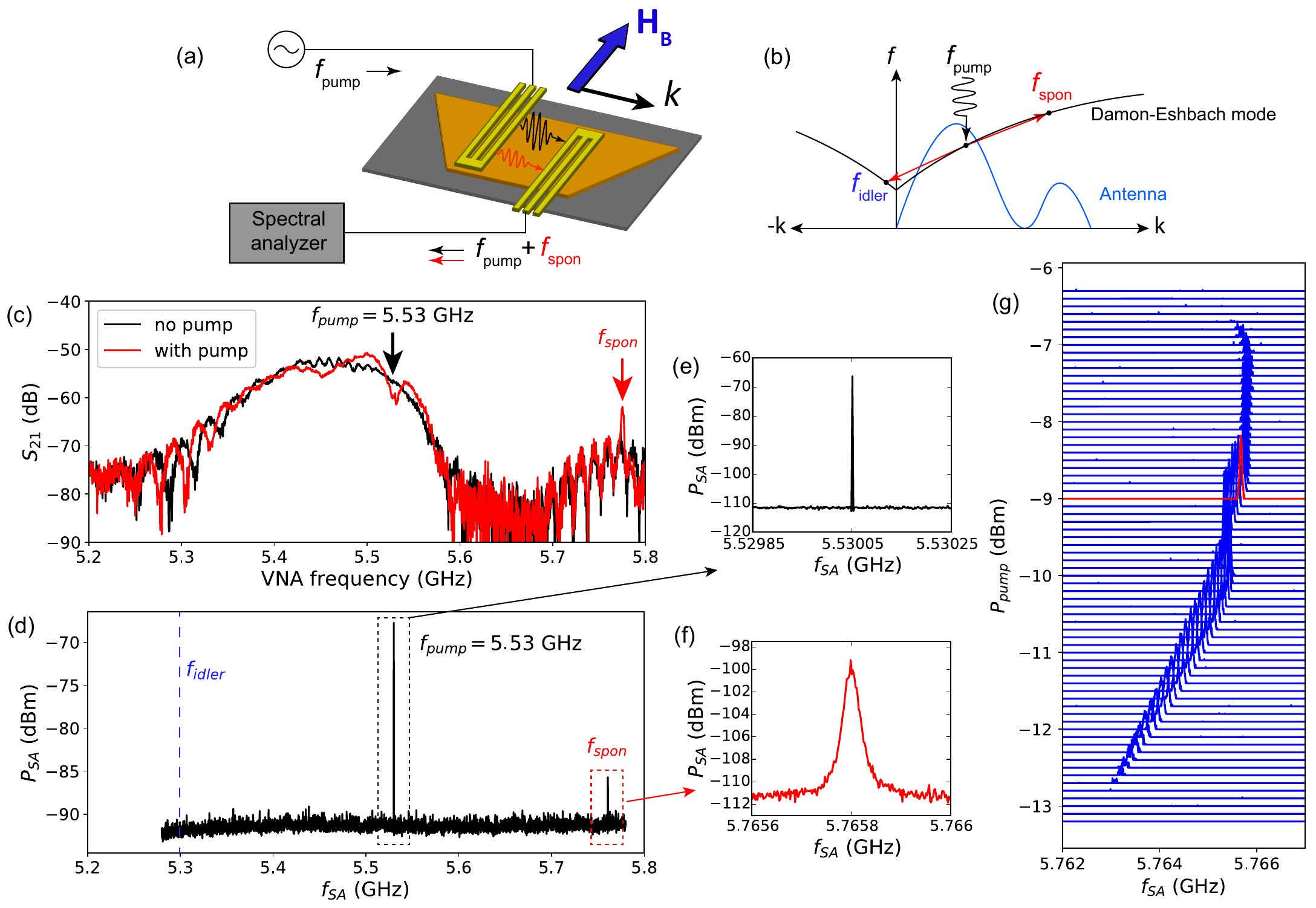}
 \caption{(a) Schematic of spontaneous mode measurements of a YIG delay line using a spectrum analyzer. The magnetic field is set at $\mu_0H_B=0.11$ T. (b) Illustration of 4-wave mixing using Damon-Eshbach magnetostatic modes, where two $f_\text{pump}$ magnons is converted to one $f_\text{spon}$ magnon and one $f_\text{idler}$  magnon. The antenna efficiency function is also plotted with a blue curve showing the first and third harmonics. (c) Comparison of VNA measurements with pump on and off at $f_\text{pump}=5.53$ GHz and $P_{pump}=-9$ dBm. The red arrow indicates the spontaneous mode. (d) SA measurement of spontaneous mode at $f_\text{pump}=5.53$ GHz and $P_{pump}=-9$ dBm, with a resolution bandwidth (RBW) of 20 kHz. (e-f) Zoom-in measurements of (e) the pump signal $f_\text{pump}$ and (f) the spontaneous mode signal $f_\text{spon}$ with a RBW of 200 Hz. (g) Evolution of the spontaneous mode peak at different $P_{pump}$. The red trace corresponds to the data shown in (f).}
 \label{fig1}
\end{figure*}

Fig. \ref{fig1}(a) shows the configuration of the YIG delay line with two parallel coplanar waveguides (CPW) patterned on top of a 100-nm-thick YIG stripe. The CPW electrode widths and the gaps between adjacent electrodes are 500 nm. The geometry determines a magnon transmission band with a maximal coupling efficiency to 2-$\mu$m-wavelength spin waves but along with a broad bandwidth as illustrated in Fig. \ref{fig1}(b) and experimentally measured in Fig. \ref{fig1}(c) using a vector network analyzer (VNA). The magnetic field $H_B$ is parallel to the CPWs and perpendicular to the spin wave wave vector $k$, as labeled in Fig. \ref{fig1}(a), so that the Damon-Eshbach (DE) propagating magnon mode is excited. Note that under this configuration the spin wave excitation is nonreciprocal, and we take the configuration with a high transmission; see Ref. \cite{LiAPL23} for more details about spin wave nonreciprocity from the same device. At $\mu_0H_B=0.11$ T, as a microwave pump is applied within the magnon transmission band and with a proper power, e.g., $f_\text{pump}=5.53$ GHz at $P_\text{pump}=-9$ dBm, two output peaks can be measured using a spectrum analyzer (SA) as shown in Fig. \ref{fig1}(d): one at the pump frequency [$f_\text{pump}$, Fig. \ref{fig1}(e)], and the other at 5.766 GHz [$f_\text{spon}$, Fig. \ref{fig1}(f)]. The second peak shows a sharp linewidth of $\Delta f=23.5$ kHz, corresponding to a quality factor of 245,000. Its existence can be also observed in the VNA measurement in the presence of an additional pump signal [red curve in Fig. \ref{fig1}(c)]. The excitation of the second peak happens within a limited range of $P_\text{pump}$, from -12.7 dBm to -6.8 dBm as shown in Fig. \ref{fig1}(g), with the lower bound slightly above the Suhl instability threshold of $P_\text{pump}=-14$ dBm as measured by the VNA \cite{supplement}. Since there is no integer or fractional integer ratio between $f_\text{pump}$ and $f_\text{spon}$, the dominating magnon scattering process is 4WM \cite{PirroPRL14}, where two pump magnons at $f_\text{pump}$ are scattered to a spontaneous magnon mode at $f_\text{spon}$ and another idler magnon mode at $f_\text{idler}=2f_\text{pump}-f_\text{spon}$. The integrated power measured by SA is up to 5\% of the measured $f_\text{pump}$ output at $P_\text{pump}=-9$ dBm, showing a high energy conversion efficiency of the 4WM process; see the Supplemental Materials for calculation \cite{supplement}. The upper bound happens when the pump signal induces a large distortion of the spin wave dispersion curve in Fig. \ref{fig1}(b) such that the 4-wave mixing condition is no longer satisfied. We note that $f_\text{idler}$ does not appear in the SA spectrum in Fig. \ref{fig1}(d). We will show evidence of $f_\text{idler}$ mode later in the paper.

\begin{figure}[htb]
 \centering
 \includegraphics[width=3.0 in]{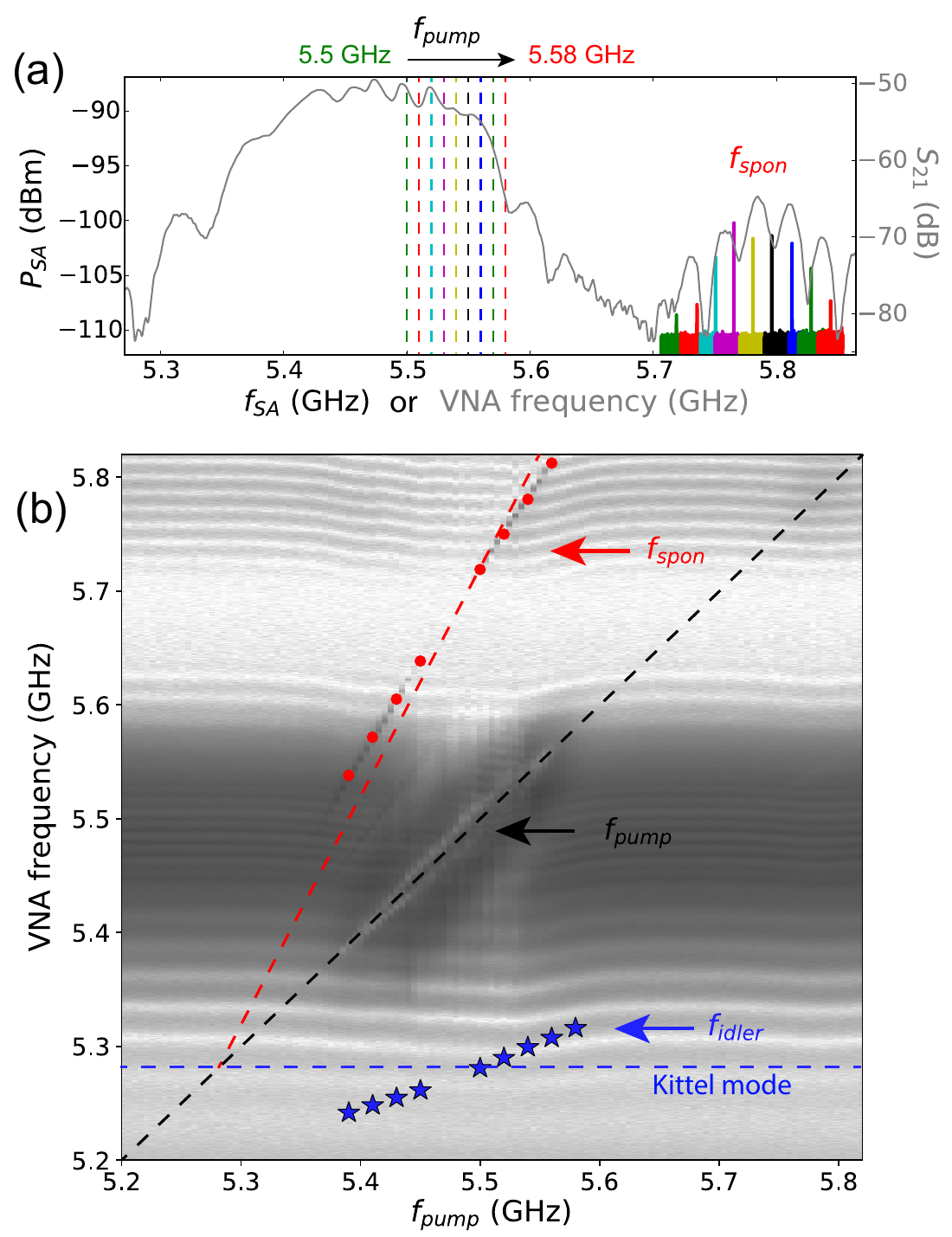}
 \caption{(a) Evolution of the spontaneous mode peaks (solid curves) at different $f_\text{pump}$ (dashed lines, from 5.5 GHz to 5.58 GHz), with the peak amplitudes following the VNA-measured magnon transmission band (gray curve). (b) VNA pump-probe measurements as a function of $f_\text{pump}$ at $P_\text{pump}=-9$ dBm. Red circles label SA-measured $f_\text{spon}$ as a function of $f_\text{pump}$. The black dashed line denotes $f_\text{pump}$. The blue dashed line denotes $f_\text{K}$. The red dashed line denotes $2f_\text{pump}-f_\text{K}$.}
 \label{fig2}
\end{figure}

To explore the relationship between the pump and the spontaneous mode, we show the evolution of $f_\text{spon}$ with $f_\text{pump}$ in Fig. \ref{fig2}. By changing $f_\text{pump}$ from 5.50 to 5.58 GHz, the spontaneous mode mode can be detected and $f_\text{spon}$ changes from 5.7187 to 5.8438 GHz, \textcolor{red}{as shown in \ref{fig2}(a)}, showing a wide range of frequency tunability. The evolution of $f_\text{spon}$ matches well with the additional peaks in VNA pump-probe measurements as shown in Fig. \ref{fig2}(b). In addition, we plot the expected $f_\text{idler}=2f_\text{pump}-f_\text{spon}$ in blue stars. The location of $f_\text{idler}$ is close to the Kittel mode ($f_\text{K}$, $k=0$) as shown the in blue dashed line, and $f_\text{spon}$ is close to $2f_\text{pump}-f_\text{K}$ as shown in the red dashed line. Note that because the spin wave wavelengths are above 1 $\mu$m, the DE mode has a quasi-linear $f$-$k$ dispersion curve, Thus, $2f_\text{pump}-f_\text{K}$ is a good approximation of the spin wave mode with twice the wavenumber as the $f_\text{pump}$ mode. The comparisons show that the 4WM process converts two $f_\text{pump}$ magnons with wavenumber $k$ to one quasi-Kittel mode with nearly zero wavenumber and another spin wave mode with a wavenumber close to $2k$. We highlight that the 4WM-generated spontaneous mode can be robustly observed in a broad frequency range. The upper frequency bound is limited by the CPW antenna bandwidth. The lower frequency bound is cause by the reduced pump efficiency as $f_\text{pump}$ starts to move out of the antenna transmission band.

\subsection{Phase locking}

\begin{figure*}[htb]
 \centering
 \includegraphics[width=6.0 in]{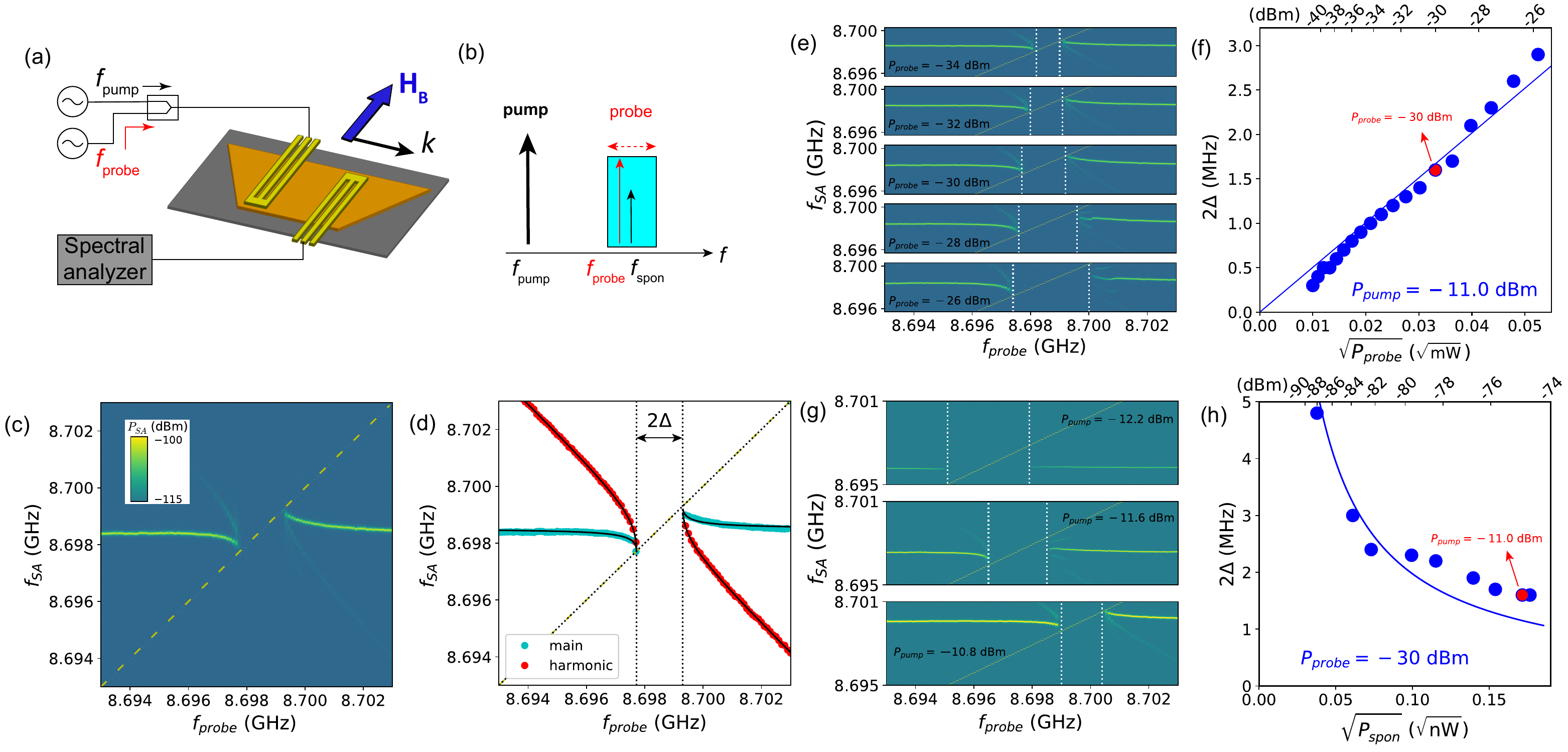}
 \caption{(a) Schematic of phase-locking measurement with pump and probe tones mixed as the input. The magnetic field is set at $\mu_0H_B=0.21$ T and the pump frequency is set to $f_\text{pump}=8.51$ GHz. (b) Illustration of the frequency relationship among $f_\text{pump}$, $f_\text{probe}$ and $f_\text{spon}$. (c) Phase locking of the spontaneous mode as a function of $f_\text{probe}$, with $P_\text{pump}=-11.0$ dBm and $P_\text{probe}=-30$ dBm. (d) Fitting curves plotted on top of the extracted peak positions for the main mode (cyan) and the first harmonic mode (red). (e) Phase locking spectra measured at $P_{probe}=-34$, -32, -30, -28 and -26 dBm, with $P_{pump}=-11.0$ dBm. (f) Extracted phase-locking bandwidth $\Delta$ as a function of $\sqrt{P_\text{probe}}$. (g) Phase locking spectra measured at $P_{pump}=-12.2$, -11.6 and -10.8 dBm, with $P_{probe}=-30$ dBm. (h) Extracted phase-locking bandwidth as a function of square root of spontaneous mode power.}
 \label{fig3}
\end{figure*}

To explore the nature of the spontaneous oscillation dynamics, we perform phase locking measurements, where a probe tone is mixed with the pump tone using a power splitter [Fig. \ref{fig3}(a)]. In order to study their interaction, we sweep the probe frequency $f_\text{probe}$ across the spontaneous mode frequency $f_\text{spon}$, as illustrated in Fig. \ref{fig3}(b). Fig. \ref{fig3}(c) shows the phase locking data of the spontaneous mode to the probe tone. The probe signal is marked by the diagonal dashed line. A clear mode attraction is observed along with a harmonic mode at $2f_\text{spon}-f_\text{probe}$. The peak positions of the spontaneous mode and the harmonic mode are plotted in Fig. \ref{fig3}(d) in cyan and red circles, respectively. They can be fitted to the model of a phase auto-oscillator \cite{Adler46,SlavinIEEE09}:
\begin{equation}\label{eq01}
  f = {f_\text{spon}+f_\text{probe} \over 2} - \sqrt{\left({f_\text{spon}-f_\text{probe} \over 2}\right)^2-\Delta^2}
\end{equation}
where $\Delta$ is the phase-locking bandwidth. The harmonic mode in Figs. \ref{fig3}(c) can be fitted by $f_\text{harmonic}=2f_\text{spon}-f_\text{probe}$, with the fitting curve also shown in Figs. \ref{fig3}(d). In addition, $\Delta$ is predicted to be proportional to the probe microwave amplitude and inversely proportional to the spontaneous mode amplitude \cite{SlavinIEEE09}:
\begin{equation}\label{eq02}
  \Delta \sim \sqrt{P_\text{probe}} / \sqrt{P_\text{spon}}
\end{equation}
Here $P_\text{probe}$ is the probe microwave power \textcolor{red}{and $P_\text{spon}$ is the power of the measured spontaneous spin wave mode}. Fig. \ref{fig3}(e) and (g) show the evolution of phase locking at different $P_\text{probe}$ and $P_\text{spon}$. Note that $P_\text{spon}$ can be tuned by changing $P_\text{pump}$, as exemplified in Fig. \ref{fig1}(g); see the Supplemental Materials \cite{supplement} for the evolution of $P_\text{spon}$ at $\mu_0H_B=0.21$ T. The extracted $\Delta$ indeed follows the linear power dependence of $\sqrt{P_\text{probe}}$ and $1/\sqrt{P_\text{spon}}$ dependence, as shown by the nice match to the fitting curves in Figs. \ref{fig3}(f) and (h), respectively. The results show that the 4WM-generated magnon mode behaves as an spin-torque auto-oscillator and can adapt its phase to an external perturbation. However, we point out that this phase adaptibility comes from the parametric instability created by the pumping spin wave.

\subsection{Search for idler mode}

\begin{figure*}[htb]
 \centering
 \includegraphics[width=6.0 in]{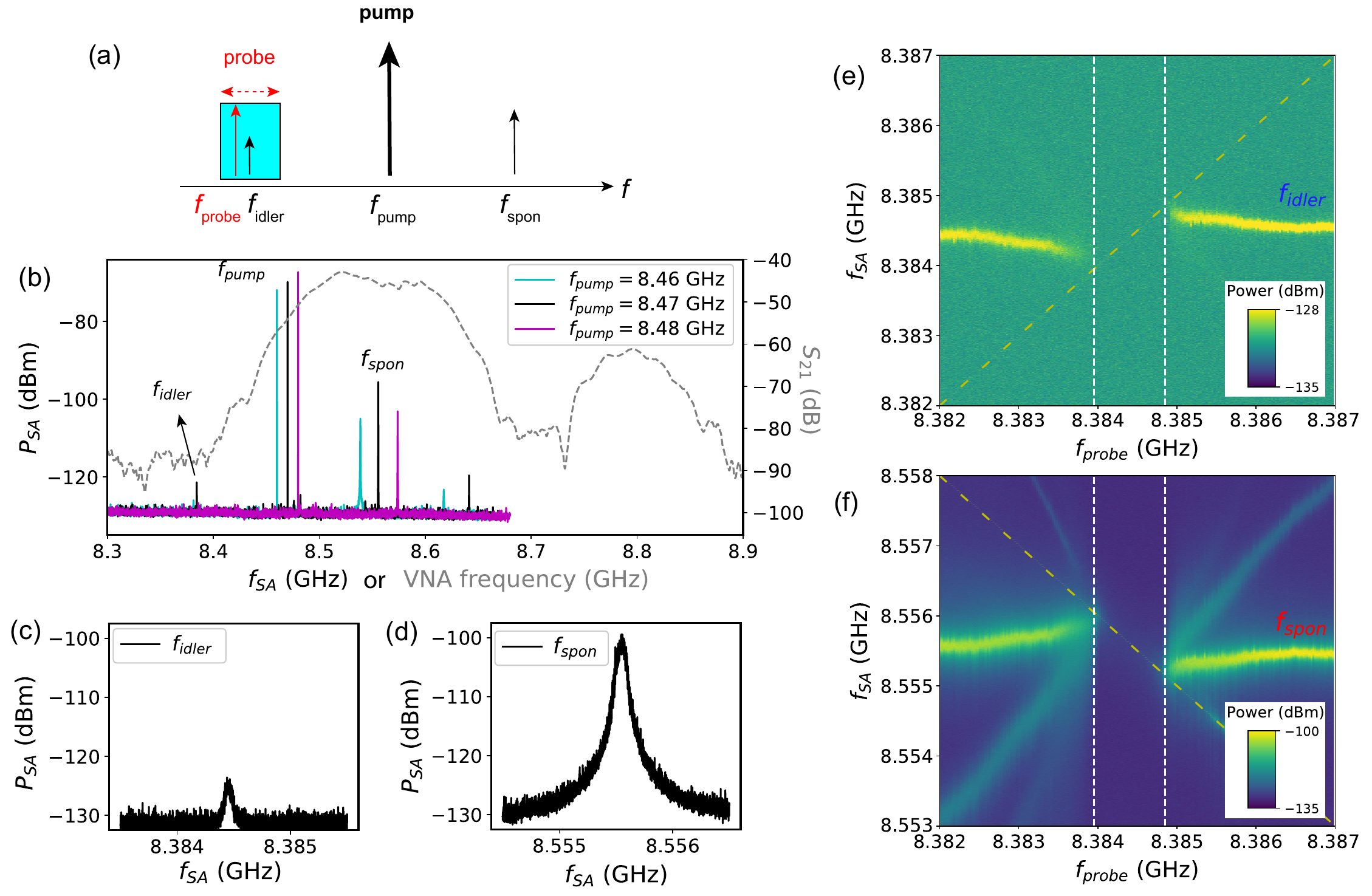}
 \caption{(a) Illustration of the frequency relationship to search for the idler mode ($f_{idler}$) with a probe tone ($f_{probe}$). (b) Measurements of the spontaneous and idler modes for $\mu_0H_B=0.21$ T at different $f_{pump}$, with $f_{pump}=8.46$ GHz showing the idler mode. (c-d) Comparison of (c) the idler mode and (d) the main spontenous mode at $f_{pump}=8.46$ GHz. (e-f) Phase locking of (e) the idler mode and (f) the main spontenous mode by applying $f_{probe}$ around $f_{idler}$ at $P_{probe}=-22$ dBm. The yellow diagonal dashed line shows the probe signal $f_{probe}$ in (e) and the mixing signal $2f_{pump}-f_{probe}$ in (f). The vertical white dashed lines show the phase-locking bandwidth.}
 \label{fig4}
\end{figure*}

The use of phase locking can help to identify the role of the idler mode. Under certain conditions, the idler mode becomes visible in the SA measurements. As shown in Fig. \ref{fig4}(b), at $\mu_0H_B=0.21$ T and $f_\text{pump}=8.47$ GHz, the idler mode appears at $f_\text{idler}=8.3845$ GHz [Fig. \ref{fig4}(c)], which satisfies the energy conservation condition $2f_\text{pump}=f_\text{spon}+f_\text{idler}$, with the measured spontaneous mode at $f_\text{spon}=8.5555$ GHz [Fig. \ref{fig4}(d)]. The idler mode is not observed at $f_\text{pump}=8.46$ and 8.48 GHz, most likely due to the reduced amplitude of the spontaneous mode and decreased 4WM efficiency. Note that higher order harmonics can also be measured similar to the reported spin-wave frequency combs \cite{HulaAPL22}, but in our case the harmonics are created by a single microwave pump. To further investigate the relation between the spontaneous mode and the idler mode, we sweep the probe frequency across $f_\text{idler}$, as illustrated in Fig. \ref{fig4}(a). In addition to the phase-locking of the idler mode to the probe [Fig. \ref{fig4}(e)], the spontaneous mode is also found to be parametrically phase-locked to the probe with the same phase-locking bandwidth [Fig. \ref{fig4}(f)]. We also observe that within the $f_\text{probe}$ scanning window, the drifts of $f_\text{spon}$ and $f_\text{idler}$ are always compensated such that $f_\text{spon}+f_\text{idler}=2f_\text{pump}$ is satisfied. This confirms the role of the idler mode and the strict energy conservation in the 4WM process.

\subsection{Parametric amplification}

With clear experimental evidence of magnon 4WM process, we further demonstrate its potential for magnon parametric amplification. Similar mechanisms have been investigated in Josephson-junction-based parametric amplifiers (JPAs) \cite{YamamotoAPL08,LehnertNPhys08,SiddiqiScience15} and kinetic-inductance parametric amplifiers (KIPAs) \cite{AnferovPRApplied20,XuPRXQ23,KhalifaPRApplied23}. Here, we demonstrate the magnonic version by leveraging the unique nonlinearity of magnonics. In conventional 4WM-based parametric amplifiers \cite{AnferovPRApplied20}, both $f_\text{pump}$ and $f_\text{idler}$ are needed for amplifying a signal at $2f_\text{pump}-f_\text{idler}$. In contrast, our magnon parametric amplification schematic requires only a single $f_\text{pump}$ tone to activate the 4WM process, thereby eliminating the complication of designing two pump inputs.

\begin{figure*}[htb]
 \centering
 \includegraphics[width=6.0 in]{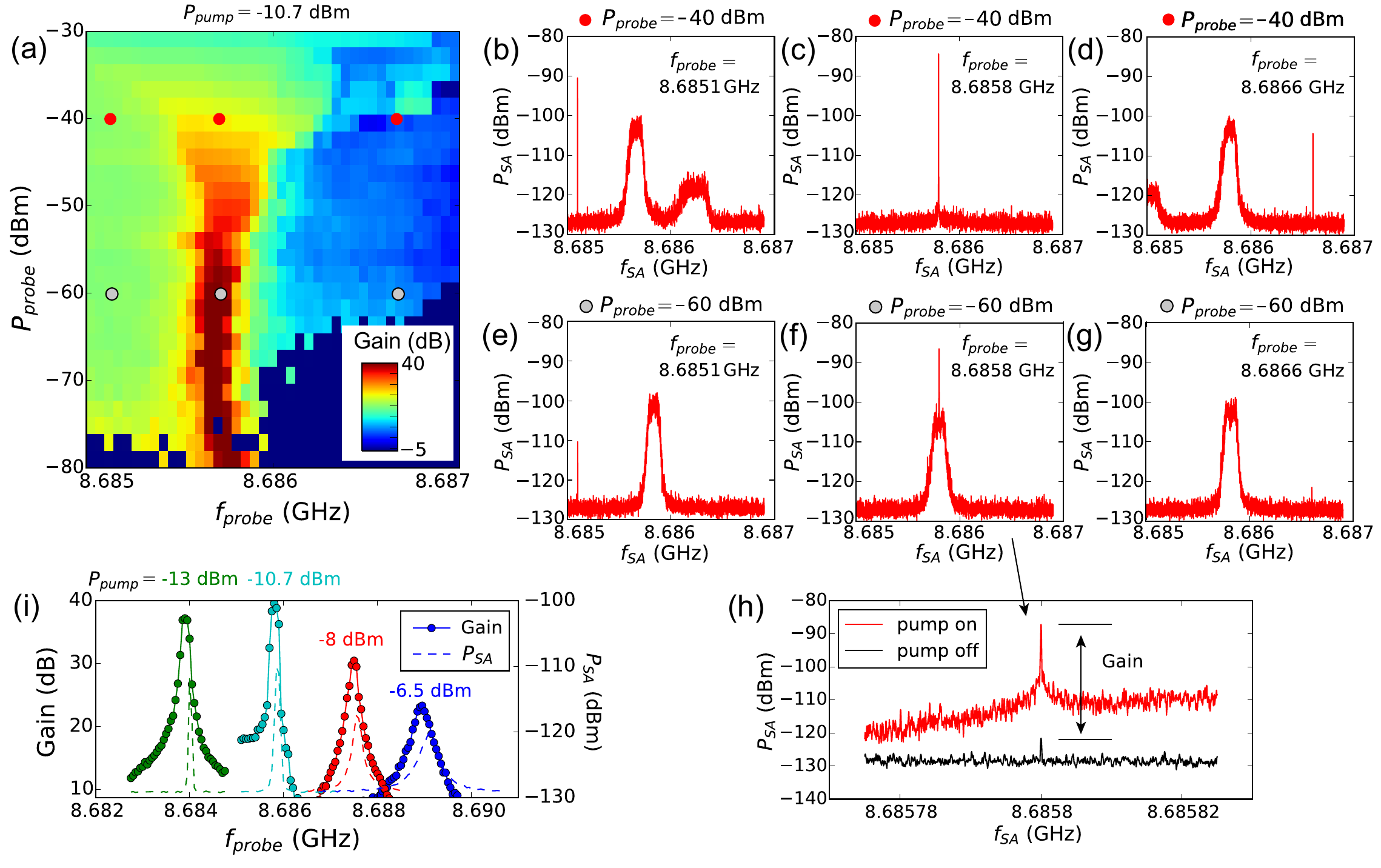}
 \caption{(a) Gain map of magnonic parametric amplification at $\mu_0H_B=0.21$ T, $f_{pump}=8.51$ GHz and $P_{pump}=-10.7$ dBm. (b-d) Spectral evolution of the mode interaction in the \textit{phase-locking} regime at $P_{probe}=-40$ dBm. (e-g) Spectral evolution of the probe signal in the \textit{constant gain} regime at $P_{probe}=-60$ dBm. (h) Comparison of probe tone lineshape with pump on and off at $P_{probe}=-60$ dBm. (i) Comparison of gain (solid curve and circles) and spontaneous mode lineshape (dashed curve) at different $P_{pump}$, measured at $P_{probe}=-60$ dBm.}
 \label{fig5}
\end{figure*}

Following the configuration shown in Fig. \ref{fig3}(a), we investigate the interaction between the probe tone and the spontaneous mode as a function of $P_\text{probe}$. Fig. \ref{fig5} (a) displays the gain of the probe tone, defined as the ratio of output power at $f_\text{probe}$ with and without the pump tone. Two distinct regimes are observed. The first is the \textit{phase-locking} regime, which occurs when $P_\text{probe} \geq -50$ dBm. In this regime, the probe signal is amplified when $f_\text{probe}$ falls within the phase-locking bandwidth. Representative output lineshapes at $P_\text{probe}=-40$ dBm are shown in Fig. \ref{fig5}(b-d). During phase-locking, as illustrated in Fig. \ref{fig5}(c), the output microwave signal at $f_\text{probe}$ contains contributions from both the probe tone and the spontaneous mode, resulting in a linear increment of output power. As a result, the gain decreases with increasing $P_\text{probe}$, while the phase-locking bandwidth correspondingly broadens. The second regime is the \textit{constant gain} regime, which is shown in Fig. \ref{fig5}(a) when $P_\text{probe} \leq -50$ dBm. As shown in Fig. \ref{fig5}(e-g) at $P_\text{probe}=-60$ dBm, when $f_\text{probe}$ is swept across the spontaneous mode peak, the probe signal climbs onto the peak profile with minimal impact on it. This is a nonlinear magnon mixing process and is sharply different from the phase locking regime. A gain up to 40 dB is obtained when $f_\text{probe}$ matches with the maximal spontaneous mode power output, with the zoom-in lineshape comparison shown in Fig. \ref{fig5}(h) at $f_\text{probe}=8.6858$ GHz. The dark blue area in Fig. \ref{fig5}(g) denotes the condition where the probe signal after nonlinear magnon mixing is below the noise background and is thus unmeasurable.

In Fig. \ref{fig5}(i), we plot the gain of magnon parametric amplification (solid curves and circles) at different $P_\text{pump}$, along with the spontaneous mode profiles (dashed curves) measured for each $P_\text{pump}$. The gain profiles are highly correlated with the spontaneous mode outputs in terms of peak positions, heights and linewidths. This shows that the gain is determined by the power spectral density of the spontaneous mode at the specific frequency. We point out that the amplification mechanism is different from cavity-based parametric amplifiers \cite{YurkeJLT06,ClerkRMP10}, which only work below the threshold of auto-oscillation. The spontaneous-mode-induced magnon parametric amplification is a unique feature of nonlinear magnonics.

\subsection{Discussion}

Our results demonstrate, for the first time, that propagating magnons can undergo phase locking to an external stimulus. This new concept is important for developing novel spin wave logic functionalities based on synchronization physics \cite{Dieny20}, and can benefit from the intrinsic spin wave nonreciprocity \cite{LiAPL23} for defining unidirectional information flow. The linewidth of the $f_\text{spon}$ mode is determined by the thermal noise of the YIG delay line device, which expands the 4WM selection rule to a finite frequency range. Compared with spin torque nano-oscillators (STNOs), the linewidth in our work is orders or magnitude narrower. This is primarily due to the much larger effective volume of the YIG thin film coupled to the 30-$\mu$-long CPW antenna, which suppresses thermal fluctuations as compared to nano-junctions required for high current densities. Additionally, the use of parametric pumping enables highly efficient energy conversion from the pump to the spontaneous mode, achieving an energy conversion efficiency up to 5\%, as shown in the Supplemental Materials \cite{supplement}. This approach reduces Joule heating and will significantly improve device lifetime, addressing a major problem in the application of STNOs.

Our demonstration of parametric amplification of magnons reveals unique physics in the nonlinear interaction between an external microwave probe and phase-autonomous propagating magnons, yielding a gain of up to 40 dB in the weak-power limit. This process is different from previous reports of magnon parametric amplification, which mainly rely on 3WM process \cite{BaoAPL08,BracherAPL14,BracherAPL14_2,VerbaAPL18,Bracher17}. In those studies, 3WM generates magnons at half the pump frequency which maintain fixed phase relation with the pump tone. This means the half-frequency magnon mode will not be able to adapt its phase to another external probe tone, and will not lead to parametric amplification with phase-sensitive nonlinear interaction as reported in our work. Furthermore, compared with the recent reports of spin-torque amplification of spin waves \cite{RezendeAPL11,NavabiPRApplied19,BreitbachPRL23,MerboucheNComm24} which use DC current to generate spin-orbit antidamping torque, the parametric spin wave amplification scheme allows for low heat generation, which are key advantages for building functional magnonic networks \cite{WangPRApplied24}.

We also note several reports on the generation of phase-autonomous magnon modes through 4WM \cite{PirroPRL14,MohseniPRL21,AnNComm24,GeilenAPR25} and 3WM \cite{SchultheissPRL19,KorberPRL20}. Notably, a similar coherent mode generated by 4WM has been electrically detected in a YIG delay line. The main difference of our work is that we identify a routine for consistent and reliable 4WM with strong power output, narrow linewidth, and broad frequency tunability up to 300 MHz by changing $f_\text{pump}$ and from 5 to 8 GHz by changing $H_B$. The tuning range of the spontaneous mode has no theoretical limit and can be extended to any value supported by $H_B$ and the magnon transmission band defined by the CPW geometry. This provides a good foundation of bandwidth for future spin wave information processing based on nonlinear magnonics.

\section{Method}

\subsection{Sample fabrication}

The YIG delay line is fabricated using a 100-nm-thick YIG film grown on gadolinium gallium garnet (GGG) substrate by liquid phase epitaxy, purchased from Matesy GmbH. The continuous YIG film is patterned to a trapezoid-shape waveguide using Ar$^{+}$ ion milling with a Cr hard mask. The mask is then removed using Cr etchant from Sigma-Aldrich. A pair of Ti(5 nm)/Au(50 nm) electrodes is subsequently fabricated on the YIG pattern using E-beam lithography and evaporation growth. Each CPW consists of three Ti(5 nm)/Au(50 nm) electrodes with a lateral width and gap of 500 nm, and a length of 30 $\mu$. The distance between the signal lines of the two CPWs is 10 $\mu$m. Another layer of Ti(10 nm)/Au(200 nm) electrodes is fabricated using photolithography in order to extend the narrow CPWs on YIG to the GGG substrate across the 100-nm YIG pattern step for probe station experiments.

\subsection{Magnon spontaneous mode measurements}

The magnon transmission and spontaneous mode experiments are conducted using a Lakeshore microwave probe station with an electromagnet at room temperature. Two types of electrical measurements are used to characterize the magnon excitations: VNA and SA. The VNA is used for measuring the magnon transmission band ($S_{21}$) without and with a microwave pump, with a constant probe power of -30 dBm and a resolution bandwidth of 100 Hz. The pump signal is generated from a second microwave synthesizer and mixed with the probe signal to the input coaxial line using a power splitter. The SA is used for measuring the 4WM signals, including the pump, the spontaneous mode, the idler mode and the probe signal. The resolution bandwidth varies from 200 Hz for rough scans to 20 Hz for fine scans, along with an average number of 10 measurements for each SA trace. The phase-locking and parametric amplification experiments are measured by SA using the same power splitter connection as for the VNA measurements. The magnetic field is applied parallel to the CPW so that the magnon transmission between the two CPWs follows the Damon-Eshbach configuration. Since the Damon-Eshbach spin wave excitation is nonreciprocal, the field direction is such that a higher magnon transmission band is obtained between the two CPW antennas with sub-micron electrode widths \cite{LiAPL23}.

\section{Data availability}

The datasets generated and/or analyzed during the current study are available from the corresponding authors on reasonable request.

\section{Ethics declarations}

The authors declare no competing interests.

\section{Acknowledgement}

We thank Helmut Schultheiss, Gr\'{e}goire de Loubens, Andrei Slavin, Vasyl Tyberkevych and Wei Zhang for helpful discussion. All the experimental works, including sample preparation, microwave characterization, and data analysis, are supported by the U.S. DOE, Office of Science, Basic Energy Sciences, Materials Sciences and Engineering Division under contract No. DE-SC0022060. The lithographic patterning and fabrication of the YIG delay line performed at the Center for Nanoscale Materials, a U.S. Department of Energy Office of Science User Facility, was supported by the U.S. DOE, Office of Basic Energy Sciences, under Contract No. DE-AC02-06CH11357. Contribution from C.A. and D.A.B. to the theoretical calculation of magnon band structure and 4WM selection rule is supported by the U.S. National Science Foundation (NSF) under Grant No. DMR-2338060. Contributions from J. L. and C. L. C. to the concepts and potential applications of magnon parameteric amplification is supported by the U.S. Department of Energy, Office of Science, National Quantum Information Science Research Centers. C. K. acknowledge support from Argonne Laboratory Directed Research and Development (LDRD) project.

%

\end{document}